\documentclass[preprint,showpacs,preprintnumbers,amsmath,amssymb,nofootinbib]{revtex4}
\usepackage{bbm}
\usepackage{amsfonts}
\usepackage{booktabs}
\usepackage{mathrsfs}
\usepackage{epsfig}
\usepackage{graphicx}
\usepackage{dcolumn}
\usepackage{bm}
\usepackage{amsmath}
\usepackage{slashed}
\usepackage{subfigure}
\usepackage{color, soul}
\let\jnfont=\rm
\def\NPB#1,{{\jnfont Nucl.\ Phys.\ B }{\bf #1},}
\def\PLB#1,{{\jnfont Phys.\ Lett.\ B }{\bf #1},}
\def\EPJC#1,{{\jnfont Eur.\ Phys.\ Jour.\ C }{\bf #1},}
\def\PRD#1,{{\jnfont Phys.\ Rev.\ D }{\bf #1},}
\def\PRL#1,{{\jnfont Phys.\ Rev.\ Lett.\ }{\bf #1},}
\def\MPLA#1,{{\jnfont Mod.\ Phys.\ Lett.\ A }{\bf #1},}
\def\JPG#1,{{\jnfont J.\ Phys.\ G}{\bf #1},}
\def\CTP#1,{{\jnfont Commun.\ Theor.\ Phys.\ }{\bf #1},}
\def\ZPC#1,{{\jnfont Z.\ Phys.\ C }{\bf #1},}
\def\JHEP#1,{{\jnfont JHEP \ }{\bf #1},}
\def\Rv{\not{\hbox{\kern-1pt $R$}}}
\def\p{\not{\hbox{\kern-3pt $p$}}}

\newcommand{\notE}{\ \hbox{{$E$}\kern-.60em\hbox{/}}}
\newcommand{\notp}{\ \hbox{{$p$}\kern-.43em\hbox{/}}}

\def\D0{\mbox{D\O}}
\def\nn{\nonumber}

\begin{document}

\title{Higgs pair production with SUSY QCD correction:\\
revisited under current experimental constraints}

\author{Chengcheng Han$^1$, Xuanting Ji$^{1,2}$, Lei Wu$^{3}$, Peiwen Wu$^1$, Jin Min
Yang$^1$}

\affiliation{\vspace{0.5cm} $^1$ State Key Laboratory of Theoretical
Physics,
Institute of Theoretical Physics, Academia Sinica, Beijing 100190, China\\
$^2$ Institute of Theoretical Physics, College of Applied Science,
Beijing University of Technology, Beijing 100124, China\\
$^3$ ARC Centre of Excellence for Particle Physics at the Terascale,
School of Physics, The University of Sydney, NSW 2006, Australia}

\begin{abstract}

We consider the current experimental constraints on the parameter
space of the MSSM and NMSSM. Then in the allowed parameter space we
examine the Higgs pair production at the 14 TeV LHC via $b\bar{b}\to
hh$ ($h$ is the 125 GeV SM-like Higg boson) with one-loop SUSY QCD
correction and compare it with the production via $gg\to hh$. We
obtain the following observations: (i) For the MSSM the production
rate of $b\bar{b} \to hh$ can reach 50 fb and thus can be
competitive with $gg \to hh$, while for the NMSSM $b\bar{b} \to hh$
has a much smaller rate than $gg \to hh$ due to the suppression of
the $hb\bar{b}$ coupling; (ii) The
SUSY-QCD correction to $b\bar{b} \to hh$ is sizable, which can reach
$45\%$ for the MSSM and $15\%$ for the NMSSM within the $1\sigma$
region of the Higgs data; (iii) In the heavy SUSY limit (all soft
mass parameters become heavy), the SUSY effects decouple rather
slowly from the Higgs pair production (especially the $gg\to hh$
process), which, for $M_{\rm SUSY}=5$ TeV and $m_A<1$ TeV, can
enhance the production rate by a factor of 1.5 and 1.3 for the MSSM
and NMSSM, respectively. So, the Higgs pair production may be
helpful for unraveling the effects of heavy SUSY.

\end{abstract}

\maketitle

\section{Introduction}

The discovery of a Higgs boson at around 125 GeV has been
announced by the ATLAS and CMS collaborations \cite{ATLAS-CMS}.
Up to now, the measurements of the
Higgs boson properties are in good agreement with the
Standard Model (SM) predictions except for the enhanced
diphoton rate $\sigma/\sigma_{SM}=1.65^{+0.34}_{-0.30}$ reported by
the ATLAS collaboration.
The future precise measurements will further test the SM and allow for
a probe for new physics like supersymmetry (SUSY) which is a promising
framework to accommodate such a 125 GeV Higgs boson
\cite{ellis,ellwanger,heinemeyer,carena,cao}.
Therefore, the intensive studies of the Higgs productions and
decays are very important and urgent.

Among the productions of the Higgs boson at the LHC, the pair
production is a rare process but quite important since it can be
used to measure the Higgs self-couplings \cite{Djouadi:1999rca}. On
the experimental side, the discovery potential of Higgs pair signal
at the LHC has been studied by analyzing the decay channels $hh \to
b\bar{b}\gamma\gamma/b\bar{b}\mu^{+}\mu^{-}$ \cite{obs1}. Recently,
the jet substructure technique was applied to the Higgs pair
production in the boosted final states\cite{obs2}, such as $hh \to
b\bar{b}\tau^{+}\tau^{-}/b\bar{b}W^{+}W^{-}$
\cite{Baglio:2012np,obs3,obs4}, which was found to be powerful in
observing the events at the 14 TeV LHC with 600 fb$^{-1}$ integrated
luminosity \cite{obs4}. On the theoretical side, in the SM the main
pair production mechanism is found to be the gluon fusion $gg\to hh$
via heavy quark loops \cite{hh-lo,hh-nlo}. Numerous studies have
also been performed for Higgs pair production in new physics models
\cite{np1,np2,np3,np4,np5,np6,np7,np8}. Note that although the
bottom quark annilation $b\bar{b}\to hh$ has a much smaller rate
than the gluon fusion process in the SM \cite{bbhh-sm,bbhh-ew}, it
can be significantly enhanced via the enlarged $hb\bar{b}$ coupling
in new physics models like the Minimal Supersymmetric Standard Model
(MSSM) \cite{bbhh-mssm}.

In this work, we revisit the Higgs pair production in SUSY for two
reasons. One is that the sizable SUSY-QCD correction must be
considered for $b\bar{b}\to hh$, which has been presented in the
MSSM but not in the NMSSM \cite{cao-hh,ellwanger-hh}. The other is
that the studies should be updated by using the latest experimental
constraints including the recent LHC Higgs data, the  LHCb $B_s\to
\mu^+\mu^-$ data and the Planck dark matter relic density. It is
also notable that the masses of the third generation sparticles
involved in the SUSY-QCD correction to $b\bar{b} \to hh$ have been
pushed up to a few hundred GeV by the LHC direct searches
\cite{atlascms-sb}. So the size of such a correction will be quite
different from the previous results in the literature
\cite{Dawson,bbhh-mssm}.

This paper is organized as follows. In Section II we briefly review
the Higgs sectors in the MSSM and NMSSM and give a description
of the analytic calculation of the SUSY-QCD correction.
Then in Section III we present the numerical results of Higgs pair production
at the LHC and discuss the SUSY-QCD residual effects in the heavy
sparticle limit.  Finally, we draw the conclusion in Section IV.

\section{A description of models and analytic calculations}
In the MSSM there are two complex Higgs doublets, $H_u$ and $H_d$,
which give rise to five physical Higgs bosons: two CP-even ($h,H$),
one CP-odd ($A$) and a charged pair ($H^{\pm}$). Due to the $\mu$
term appearing in the superpotential, the MSSM suffers from the
$\mu$-problem. Besides, in order to give a 125 GeV SM-like Higgs
boson, large corrections to the Higgs mass from heavy stops is
needed, which will lead to the little fine tuning problem . To
overcome these difficulties, we can go beyond the MSSM. One
alternative is the NMSSM, which introduces a singlet Higgs field. In
the NMSSM the $\mu$ term does not appear in the superpotential.
Instead, it is generated when the singlet Higgs field develops a
vev. Also, the SM-like Higgs boson gets an extra tree-level mass
from the mixing with the singlet field and thus the stops are not
necessarily heavy to push up the Higgs mass, which alleviates the
little fine-tuning problem\cite{nsusy1,nsusy2,nsusy3}. In the NMSSM
the singlet Higgs field mixes with the other two doublet scalars.
Then the Higgs sector contains seven Higgs bosons, i.e., compared
with the five Higgs bosons in the MSSM, the NMSSM contains one more
CP-even and one more CP-odd Higgs bosons.
In the following $H_{1,2}$ denote the real scalar components of $H_{d,u}$ in the MSSM
and $H_{1,2,3}$ denote the real scalar components of $H_{d,u,s}$ in the NMSSM.  $\tan\beta \equiv v_{u}/v_{d}$ is also used in our paper
(here $H_{d}$, $H_{u}$ and $H_{s}$ are the down-type, up-type and singlet Higgs fields, respectively).
One can get the mass eigenstates from the CP-even states:
\begin{eqnarray}
&&{\rm \textbf{MSSM}}: ~~
h_{i}=U_{ij} H_j~~~~(i,j= 1,2), \\
&&{\rm \textbf{NMSSM}}:
h_{i}=V_{ij} H_j~~~~ (i,j= 1,2,3)
\end{eqnarray}
where $U_{i1}^2+U_{i2}^2=1, V_{i1}^2+V_{i2}^2+V_{i3}^2=1$ and the $h_i$ is aligned by mass.
The singlet contribution is reflected
by the rotation matrix elements ${V_{i3}}$ via the formula $h_{SM} =
{V_{h_{SM}1}}H_{1} + {V_{h_{SM}2}}H_{2} +{V_{h_{SM}3}}H_{3}$
(a large ${V_{h_{SM}3}}$ means that $h_{SM}$ has a considerable singlet component).


In our calculations, we follow the simplified ACOT prescription to
deal with the $b$-quark mass \cite{Aivazis,Collins,Kramer}. By
including the QCD and SUSY-QCD effects to the bottom Yukawa
couplings, we can respectively obtain the effective $h_{i}b\bar{b}$
couplings in the MSSM
\cite{Hollik,Enberg,Pierce:1996zz,Carena:1999py,Carena:2002bb,Guasch:2003cv,Noth:2008tw,Noth:2010jy,Mihaila:2010mp}
and NMSSM\cite{Baglio:2013iia}:
\begin{eqnarray}
&& {\rm \textbf{MSSM}}:~~~~~~y_{h_{i}bb} \rightarrow \frac{g
m^{\overline{{DR}}}_b}{2M_W} \frac{U_{i1}}{\cos\beta}
\Delta_{bi}^{MSSM}~~~ (i=1,2),\\
&& {\rm \textbf{NMSSM}}: ~~~~y_{h_{i}bb} \rightarrow \frac{g
m^{\overline{{DR}}}_b}{2M_W} \frac{V_{i1}}{\cos\beta}
\Delta_{bi}^{NMSSM}~~~ (i=1,2,3)
\end{eqnarray}
where
\begin{eqnarray}
&& \Delta_{bi}^{MSSM}=\frac{1}{1+\Delta_b^{1}}\left(1+\Delta_b^{1}
\frac{U_{i2}}{U_{i1}\tan\beta}\right)  ~(i=1,2),\label{deltab}\nonumber\\
&&
\Delta_{bi}^{NMSSM}=\frac{1}{1+\Delta_b^{1}}\left[1+\Delta_b^{1}\left(
\frac{V_{i2}}{V_{i1}\tan\beta}+
\frac{V_{i3}{v_{d}}}{V_{i1}v_{s}}\right)\right]  (i=1,2,3),\label{deltab}\nonumber\\
&& \Delta_b=\frac{2\alpha_s}{3\pi} m_{\tilde g} \mu \tan\beta
I(m^2_{\tilde b_1}, m^2_{\tilde b_2}, m^2_{\tilde g})\nonumber\\
&&\Delta_b^{2}= -\frac{2\alpha_s}{3\pi} m_{\tilde g} A_b
I(m^2_{\tilde b_1}, m^2_{\tilde b_2}, m^2_{\tilde
g}),\Delta_b^{1}=\frac{\Delta_b}{1+\Delta_b^{2}}
\end{eqnarray}
Here it should be noted that due to the contribution of the singlet
field to the effective potential, an additional correction term
$\Delta_b^{1}\frac{V_{i3}{v_{d}}}{V_{i1}v_{s}}$ appears in the
NMSSM. The $v_{d}$ and $v_{s}$ are the VEVs of the Higgs fields $H_u$ and $H_d$ respectively. The
auxiliary function $I$ is defined as
\begin{eqnarray}
I(a,b,c) = -\frac{1}{(a-b)(b-c)(c-a)}(ab\ln\frac{a}{b}+b
c\ln\frac{b}{c}+c a\ln\frac{c}{a}) \,
\end{eqnarray}
. The value of $m^{\overline{{DR}}}_b$ is related to the
QCD-$\overline{\rm MS}$ mass $m_b^{\overline{\text{MS}}}$ (which is
usually taken as an input parameter \cite{E.braaten}) by
\begin{eqnarray}
 m^{\overline{{DR}}}_b (\mu_R)=m^{\overline{{MS}}}_b(\mu_R)\left[1 - \frac{\alpha_s}{3\pi} -
\frac{\alpha_s^2}{144\pi^2}(73-3n) \right],
\end{eqnarray}
where $n$ is the number of active quark flavors and
$m^{\overline{{MS}}}_b(\mu_R)$ is taken as
\begin{eqnarray}
m_b^{\overline{\text{MS}}}(\mu_R) =
\begin{cases}
U_6(\mu_R, m_t)U_5(m_t, \overline{m}_b)\overline{m}_b(\overline{m}_b) \quad & \text{for}\quad \mu_R > m_t \\
U_5(\mu_R, \overline{m}_b)\overline{m}_b(\overline{m}_b) \quad &
\text{for}\quad \mu_R \le m_t.
\end{cases}
\label{mb_evolution}
\end{eqnarray}
When $Q_2 > Q_1$, the evolution factor $U_n$ reads
\begin{eqnarray}
U_n(Q_2,Q_1)=\left(\frac{\alpha_s(Q_2)}{\alpha_s(Q_1)}\right)^{d_n}\left[1+\frac{\alpha_s(Q_1)-\alpha_s(Q_2)}{4\pi}J_n\right],
\end{eqnarray}
where
\begin{eqnarray}
d_n&=&\frac{12}{33-2n} , ~~J_n = -\frac{8982 - 504n + 40n^2}{3(33 -
2n)^2}.
\end{eqnarray}

 Since the $\Delta_b$-related
corrections have already been included into the tree-level
contribution, we need the following counter terms to subtract them
to avoid double counting in the one-loop calculations \cite{Hollik}
\begin{eqnarray}
&& {\rm \textbf{MSSM}}: ~~~~\delta \tilde{m}_b^{h_{i}} =
m^{\overline{{DR}}}_b \left(1 -
\frac{U_{i2}}{U_{i1}\tan\beta}\right)\Delta_b^{1},\quad (i=1,2),\\
&& {\rm \textbf{NMSSM}}: ~~~~\delta \tilde{m}_b^{h_{i}} =
m^{\overline{{DR}}}_b \left(1 - \frac{V_{i2}}{V_{i1}\tan\beta}-
\frac{V_{i3}{v_{d}}}{V_{i1}v_{s}}\right)\Delta_b^{1},\quad
(i=1,2,3).
\end{eqnarray}

For SUSY-QCD corrections to $b\bar{b}\to hh$, the sbottoms and gluino are
involved in the loops. The sbottom mass matrix is given by \cite{J. F. Gunion}
\begin{equation}
M_{\tilde b}^2 =\left(\begin{array}{cc}
m_{{\tilde b}_L}^2& m_bX_b^\dag\\
 m_bX_b& m_{{\tilde b}_R}^2 \end{array} \right) \ ,\label{sbottom}
\end{equation}
where
\begin{eqnarray}
m_{{\tilde b}_L}^2 &=& m_{\tilde Q}^2+m_b^2-m_Z^2(\frac{1}{2}
-\frac{1}{3}\sin^2\theta_W)\cos(2\beta) \ , \nonumber \\ m_{{\tilde
b}_R}^2 &=& m_{\tilde D}^2+m_b^2 -\frac{1}{3}m_Z^2
\sin^2\theta_W\cos(2\beta) \ ,\nonumber \\  X_b&=& A_b-\mu\tan\beta .
\end{eqnarray}
After diagonalizing Eq.(\ref{sbottom}), we can obtain the
sbottom masses $m_{\tilde b_{1,2}}$ and the mixing angle
$\theta_{\tilde{b}}$:
\begin{eqnarray}
m_{\tilde b_{1,2}}&=&\frac{1}{2}\left[ m_{{\tilde b}_L}^2+m_{{\tilde
b}_R}^2
\mp\sqrt{\left(m_{{\tilde b}_L}^2-m_{{\tilde b}_R}^2\right)^2 +4m_b^2X_b^2}\right],\nonumber\\
\tan2\theta_{\tilde{b}} &=& \frac{2m_bX_b}{m_{{\tilde
b}_L}^2-m_{{\tilde b}_R}^2} \ .
\end{eqnarray}

The Feynman diagrams for one-loop SUSY-QCD corrections to $b\bar{b}
\to hh$ has been represented in \cite{Dawson}. To preserve
supersymmetry, we adopt the dimension reduction method to regulate
the UV divergences in the gluino and squark loops. Then we use the
on-shell renormalization scheme to remove these UV divergences.

\section{Numerical studies}
\subsection{A scan of parameter space}

We use NMSSMTools \cite{NMSSMTOOLS} and LoopTools \cite{LoopTools}
to perform a random scan over the parameter space and loop
calculations. For simplicity, we assume an universal parameter $M_{L3}$ for the slepton sector
and fix all irrelevant soft parameters for first two generation of the squark sector to be 1 TeV.
We also set $M_{D3}=M_{U3}$ and $A_b =A_t$ for the third generation of the squarks.
Besides, we impose the grand unification relation of the gaugino masses,
$3 M_1/5\alpha_1 = M_2/\alpha_2 = M_3/\alpha_3$, and treat $M_1$ as an input parameter.
The parameter ranges in our scan are:
\begin{itemize}
\item[(a)] For the MSSM
\begin{eqnarray}
&&1 \le \tan\beta \le 60, ~ 100 {~\rm GeV} \le \ M_A \le 1 {~\rm
TeV}, ~ 100 {~\rm GeV} \le\mu \le 2 {~\rm TeV}  \nonumber \\&&100
{~\rm GeV} \le M_{Q3},M_{U3} \le 2 {~\rm TeV}, 100{~\rm GeV} \le M_{L3} \le 1 {~\rm TeV} \nonumber \\
&& |A_t|\le 5 {~\rm TeV},  ~50 {~\rm GeV}\le M_{1} \le 500 {~\rm
GeV}.
\end{eqnarray}
\item[(b)] For the NMSSM
\begin{eqnarray}
&&0.5 \le \lambda \le 0.7, ~ 0.1 \le \kappa \le 0.51, ~ |A_\kappa|
\le 1 {~\rm TeV} \\ \nonumber &&1 \le \tan\beta \le 10, ~ 100 {~\rm
GeV} \le \mu \le 600 {~\rm GeV}, ~ 100 {~\rm GeV} \le M_A \le 1
{~\rm TeV} \\ \nonumber &&100 {~\rm GeV} \le M_{Q3},M_{U3} \le 2
{~\rm TeV},100{~\rm GeV} \le M_{L3} \le 1 {~\rm TeV}  \\ \nonumber
&&|A_t|\le 5 {~\rm TeV}, ~50 {~\rm GeV}\le M_{1} \le 500 {~\rm GeV}.
\end{eqnarray}
\end{itemize}

In our scan we consider the following experimental constraints:
\begin{itemize}
\item[(i)] The bounds for Higgs boson from the
LEP, Tevatron and LHC experiments and require the SM-like Higgs
mass to be in the range of 123 GeV$<m_{h}<$ 127 GeV;
Here we require the surviving samples to explain the observable at
$2\sigma$ level which has an experimental central value. For the LEP
and Tevatron limits, the upper or lower bounds are implemented in
our scan. For the LHC Higgs search of
$H/A\rightarrow\tau\tau$\cite{htautau} and $H^\pm\rightarrow\tau\nu_\tau$\cite{hnvtau},
 we require the samples
to satisfy the upper limits.


\item[(ii)] The constraints from the precision electroweak data \cite{electroweak}
and flavor physics at $2\sigma$ level;
\item[(iii)] The dark matter relic density from Plank at $3\sigma$ level
and the limit of direct detection from XENON100 \cite{Darkmatter};
\item[(iv)] The explanation of muon $g-2$ at $2\sigma$ level \cite{g-2}.
\end{itemize}

\begin{figure}[htb]
\centering\leavevmode \epsfxsize=5.5in
\epsfbox{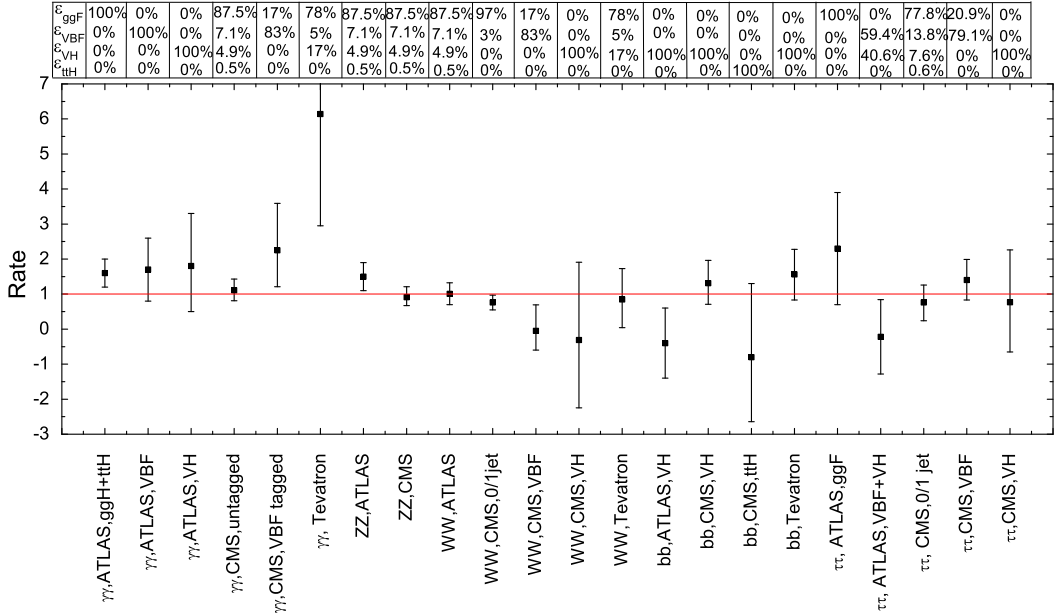}\vspace{-1cm} \caption[]{The measured signal
strength of Higgs boson with their $1\sigma$ error-bars and
selection efficiencies $\epsilon_p$ for each production mode $p$ and
decay mode at the 7+8 TeV LHC and Tevatron.} \label{fig:higgsdata}
\end{figure}
In our scan, for each experimental data which has a central value,
we require the samples to agree with the experimental data at
$2\sigma$ level, except for the dark matter relic density which is
required to agree with the measured value at $3\sigma$ level (we
made such a choice just in order to be consistent with the analysis
in the literature). For the LEP and Tevatron direct search bounds on
sparticle masses, we just require the samples to satisfy such
bounds. For the LHC Higgs search of $H/A\rightarrow\tau\tau$ and
$H^\pm\rightarrow\tau\nu_\tau$, we require the samples to satisfy
the upper limits. The scan ranges of the parameters are large, we
keep the samples survived various experimental constraints as stated
above. Besides, we further require gluino mass larger than 1 TeV to
avoid multi-jets search on SUSY\cite{multijet}.
 However, we did not impose
other LHC direct limits on sparticles for the following reasons.
First, we required the first and second generations of squarks to be
1 TeV and the gluino beyond 1 TeV. But the latest LHC search results
gave more stringent constraints on such squark and gluino mass (the
most stringent bound is for the CMSSM, which is $m_{\tilde g}> 1.7$
TeV in case of  $m_{\tilde g}\simeq m_{\tilde q}$ and $m_{\tilde g}>
1.1$ TeV in case of $m_{\tilde q}\gg m_{\tilde g}$). Actually, our
results are not sensitive to these masses. Second, the current LHC
limit is about 500-600 GeV for stop and 400-600 GeV for
sbottom\cite{stop}. However, such limits were obtained in some
simplified model or by assuming a certain decay branching ratio to
be $100\%$. In our case the stop and sbottom decays are quite
complicated, which will weaken the LHC limits. Further, for
electroweak gauginos and sleptons, the current LHC limits will also
be weakened in our case for the same reason. After that we also
require surviving samples to avoid Landau singularity at GUT scale
and we checked that all of our surviving samples satisfy
$\sqrt{\lambda^2+\kappa^2}<0.75$ in NMSSM. We note that a large
$\tan\beta$ exist in the surviving samples of the MSSM, this is
because that a 125 GeV neutral Higgs mass is guaranteed by a large
$A_t$ (which provides $X_t/M_s$ close to $\sqrt{6}$) even for
$\tan\beta$ as large as 40. As for the flavor constraints, we
projected our samples onto the $\tan\beta$ versus the charged Higgs
mass plane and found that when $\tan\beta$ increases the charged
Higgs mass grows dramatically (especially, for $\tan\beta$ close to
40, the charged Higgs mass is heavier than 700 GeV) and thus can
satisfy the flavor constraints.
For the samples surviving the above constraints (i)-(iv), we further
perform a fit by using the available Higgs data at the LHC. We
define the Higgs signal strength $\mu_i$ as
  \begin{equation}
  \begin{aligned}
     &\mu_i=\frac{\Sigma_p \sigma_p \epsilon_p^i}{\Sigma_p \sigma_p^{SM} \epsilon_p^i}
      \frac{{\rm Br}_i}{{\rm Br}_i^{SM}},
     \label{mu}
  \end{aligned}
  \end{equation}
where $p$ is the Higgs boson production mode and $i$ stands for the
measured channels by Tevatron, ATLAS and CMS collaborations. For
each production mode $p$, its contribution to the channel $i$ can be
determined by the selection efficiency
$\epsilon_p^i$ \cite{Cheung:2013kla}. We summarize all experimental
signal strength $\mu_i^{exp}$ with their $1\sigma$ error-bars and
selection efficiencies in Fig.\ref{fig:higgsdata}. We can see that
most measurement results are consistent with the SM predictions. The
CMS and ATLAS collaborations also reported their observations of the
Higgs mass $M^{exp}_h$ \cite{higgsmass}:
  \begin{equation}
  \begin{aligned}
     &M^{exp}_h=\left\{
            \begin{array}{ll}
              125.8\pm0.5\pm0.2 {\rm ~GeV} & \hbox{(${\rm CMS}~ZZ$),} \\
              125.4\pm0.5\pm0.6 {\rm ~GeV} & \hbox{(${\rm CMS}~\gamma\gamma$),} \\
              124.3\pm0.6\pm0.5 {\rm ~GeV} & \hbox{(${\rm ATLAS}~ZZ$),}\\
              126.8\pm0.2\pm0.7 {\rm ~GeV} & \hbox{(${\rm ATLAS}~\gamma\gamma)$.}
            \end{array}
          \right.
     \label{mh}
  \end{aligned}
  \end{equation}
We use the combined Higgs mass $M^{exp}_{h}=125.66 \pm 0.34$ GeV\cite{giardino}. The $\chi^2$ definition in our fit is
  \begin{equation}
  \begin{aligned}
     &\chi^2= \sum\limits_{i=1}^{22} \frac{(\mu_i -\mu_i^{exp})^2}{\sigma^2_i}+\frac{(M_h-M_h^{exp})^2}{\sigma^2_{M_h}}.
     \label{chisq}
  \end{aligned}
  \end{equation}
where $\sigma_i$ and $\sigma_{M_h}$ only denote the experimental
errors.

\subsection{The cross section of $b\bar{b} \to hh$ with SUSY-QCD correction}
We use CTEQ6L1 and CTEQ6m \cite{Pumplin} for the leading order and
SUSY-QCD calculation,  respectively. The renormalization scale
$\mu_R$ and factorization scale $\mu_F$ basically can vary between
$M_{h}/2$ and $2M_{h}$. In order to compare our results with
\cite{Dawson} where $\mu_R= \mu_F =M_{h}/2$ is assumed, we also made
this assumption in our calculation. The input parameters of the SM
are taken as \cite{PDG}
\begin{eqnarray}
&&  m_b=4.7{\rm ~GeV},~m_{t}=173.1{\rm ~GeV}, ~m_{Z}=91.19 {\rm
~GeV},\nn \\
&&\sin^{2}\theta_W=0.2228,
~\alpha_s(m_t)=0.1033,~\alpha=1/128.
\end{eqnarray}

\begin{figure}[htb]
\centering\leavevmode \epsfxsize=6in \epsfbox{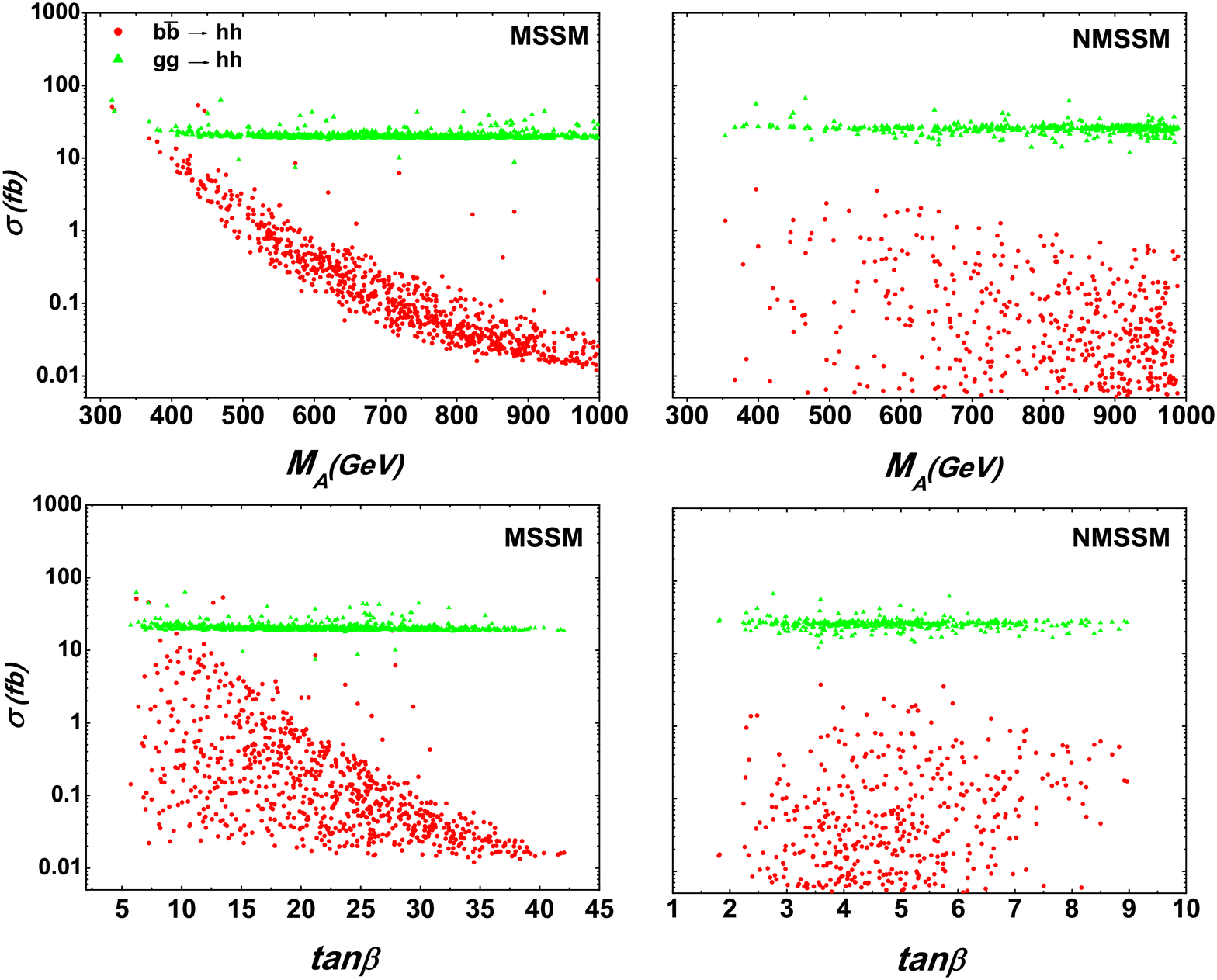}
\vspace{-0.5cm} \caption[]{The scatter plot of the parameter space
satisfying the experimental constraints (i-iv), showing the hadronic
cross sections of the SM-like Higgs pair productions via $b\bar{b}$
annihilation (with SUSY QCD correction) and $gg$ fusion versus $M_A$
and $\tan\beta$ at the 14 TeV LHC in MSSM and NMSSM.}
\label{fig:mssm1}
\end{figure}

In Fig.~\ref{fig:mssm1}, we display the parameter space satisfying
the experimental constraints (i-iv), showing the cross sections of
the SM-like Higgs pair productions via $b\bar{b}$ annihilation (with
SUSY QCD correction) and gg fusion versus $M_A$ at the 14 TeV LHC in
MSSM and NMSSM.
In this paper we aim to investigate the property of the $b\bar{b}
\to hh$ production by including the SUSY QCD corrections. For the
$gg \to hh$ production, we only calculate its cross section at
one-loop level, not including the SUSY QCD corrections due to its
small relative correction\cite{SUSYQCDgg} comparing the SUSY QCD
correction on $b\bar{b} \to hh$ process.
 We used our own codes and combined
them with Looptools to do our calculation. We checked our results
with [27] and found good agreement.

 We checked that our results agree with\cite{Dawson} for $b\bar{b} \to hh$ and
with \cite{cao-hh} for the gluon fusion process.We can see that due
to the constraints from the LHC and B-physics, such as $H/A \to
\tau^+ \tau^-$ \cite{htautau} and $B_s \to \mu^+ \mu^-$
\cite{Bsmumu}, the values of $m_A$ must be larger than about 300
GeV. In the MSSM the maximal cross section can still reach 50 fb at
14 TeV LHC, which can be competitive with $gg \to hh$. However, we
also notice that the hadronic cross section proceeding through
$b\bar{b} \to hh$ deceases when $m_A$ or $\tan\beta$ becomes large.
The reason can be understood as follows. On the one hand, for a
moderate $m_A$, the dominant contribution to $b\bar{b} \to hh$ comes
from the resonant production $b\bar{b} \to H \to hh$. With the
increase of $M_A$, the mass of $H$ gets heavy and then the
production rate of $b\bar{b} \to hh$ is suppressed. Besides, the
coupling of $hhH$ will approach to zero for a large $m_A$ and also
leads to the reduction of the cross section. On the other hand, for
a small $\tan\beta$, $H$ has a large branching ratio into a pair of
Higgses $hh$\cite{Dolan:2012ac}, for a large $\tan\beta$, the
production rate of $b\bar{b} \to H$ can be enhanced but the branch
ratio of $H \to hh$ is highly suppressed. So the total production
rate of $b\bar{b} \to hh$ will become small.The decoupling behavior
of the cross section proceeding through $gg \to hh$ can be
understood with the following considerations: To predict a 125 Gev
Higgs boson, a large $A_t$ is required, which induces a sizable SUSY
effect for the process $gg\to hh$. $M_{A}$ affects the process $gg
\to hh$ mainly through the Higgs mass $m_h$. So when we require
$m_h$ in the range of 123-127 GeV, the process $gg \to hh$ is not
sensitive to $M_{A}$. Further, since $gg \to hh$ is dominated by the
stop loops, the value of $\tan\beta$ affects this process through
the coupling $h\tilde{t_i}\tilde{t_j}$. Because this coupling is not
sensitive to $\tan\beta$ for our surviving points, our results
depend weakly on $\tan\beta$.

In NMSSM the SM-like Higgs boson $h$ with mass around 125 GeV can be
either $h_1$ or $h_2$. However, we focus on the $h=h_2$ scenario
that is more welcomed by the naturalness. From Fig.\ref{fig:mssm1}
we can see that the maximal cross section of $b\bar{b} \to hh$ can
only reach about 4 fb, which is much smaller than $gg \to hh$.
We find that the suppression of $b\bar{b} \to hh$ in NMSSM mainly has two reasons.
One is that in NMSSM the $\tan\beta$ value is around 3-5 which
 is much smaller than in MSSM which is always larger than 10.
So the $\tan\beta$ enhancement on $h_{i}b\bar{b}$ coupling is not significant in NMSSM.
The other reason is the $h_3h_2h_2$ coupling is suppressed for most surviving points
(the main reason is the cancelation of different contributions).
Besides, in the NMSSM the 125 GeV Higgs mass requires a small
$\tan\beta$ and a large $\lambda$. So the cross section of $b\bar{b}
\to hh$ can hardly enhanced by $\tan\beta$.

\begin{figure}[ht]
\includegraphics[width=14.5cm]{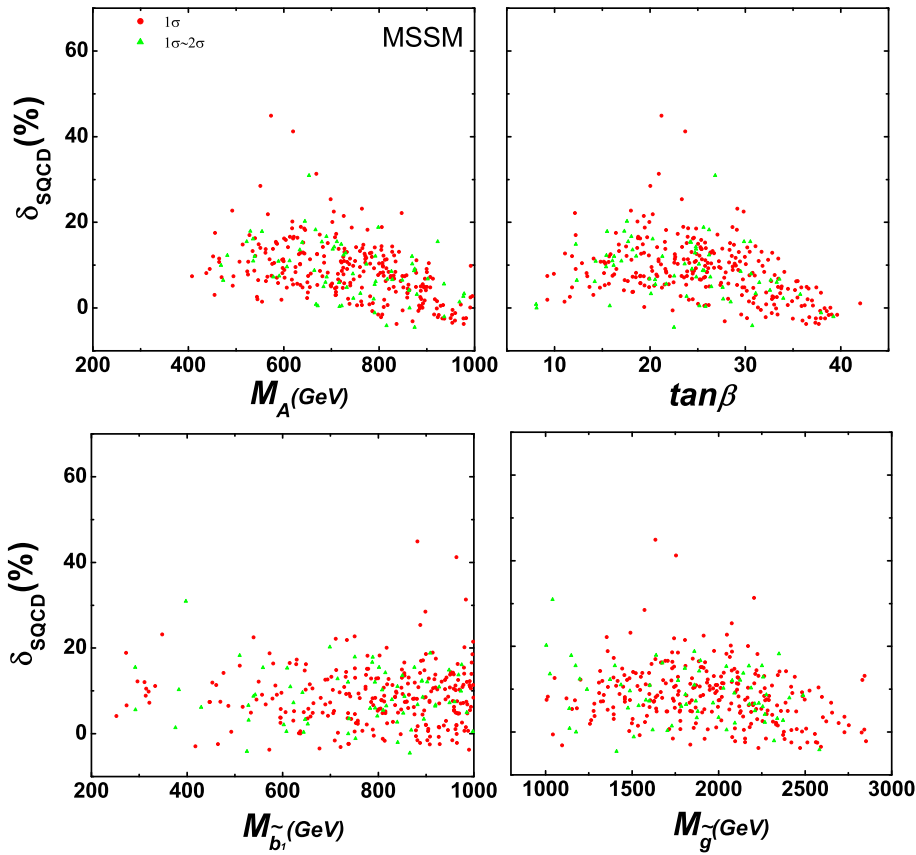}
\vspace{-0.5cm} \caption[]{Same as Fig.2, but showing the relative
SUSY-QCD correction  for the $b\bar{b} \to hh$
 in the MSSM. Here the samples
satisfying the experimental constraints (i-iv) are further
classified according to the Higgs data: within $1\sigma$ (red dots),
outside $1\sigma$ but within $2\sigma$ (green triangles).}
\label{fig:mssm2}
\end{figure}
To further investigate the influence of the Higgs data in Fig.2
on the SUSY-QCD effect in $b\bar{b} \to hh$, we define the
relative SUSY-QCD correction $\delta_{SQCD}$ as
\begin{eqnarray}
\delta_{SQCD}=\frac{\sigma_{SQCD}-\sigma_{LO}}{\sigma_{LO}} .
\end{eqnarray}
 In our calculation we use the $\alpha_{s}^{LO}$ for the
LO cross-section and $\alpha_{s}^{NLO}$ for the NLO cross-sections,
respectively. In Fig.~\ref{fig:mssm2} we show the dependence of
$\delta_{SQCD}$  for the $b\bar{b} \to hh$
 on the SUSY parameters $M_A$,
$\tan\beta$, the lightest sbottom mass ($m_{\tilde{b}_1}$) and
gluino mass ($m_{\tilde{g}}$) in the MSSM. In this figure the
samples satisfying the experimental constraints (i-iv) are further
classified according to the Higgs data: We use the $\chi^2$ and the
degree of freedom to calculate the p-value for each point and plot
the points whose p-values are larger than 0.045 (2$\sigma$) and
0.318 (1$\sigma$). The degree of freedom is 15 [23(experimental
observables)-8(free parameters)] for MSSM and 12 [23(experimental
observables)-11(free parameters)] for NMSSM.From the upper panel we
can see that a heavy $m_A$ ($>400$~GeV) and a moderate $\tan\beta$
($10\sim40$) are favored by the Higgs data and the SUSY-QCD
correction can maximally reach about $45\%$
for the samples in $1\sigma$ range. Similar to
Fig.\ref{fig:mssm1}, $\delta_{SQCD}$ decreases when $m_A$ becomes
heavy. From the lower panel we note that for heavy $m_{\tilde{b}_1}$
and $m_{\tilde{g}}$, the SUSY-QCD effects decouple slowly. This
behavior is because that the SUSY-QCD corrections depend on the
ratio of the SUSY parameters. For example, in the triangle diagrams,
the SUSY-QCD correction to the vertex $hb\bar{b}$ is proportional to
$M_{EW}^2/M_A^2$ and
$M_{EW}^2/M_{\tilde{b}}^2$\cite{Guasch:2003cv,Gao}. So only when all
the sparticles and $m_A$ are heavy, the SUSY-QCD effect can
completely decouple from the process of $b\bar{b} \to hh$.

\begin{figure}[ht]
\includegraphics[width=16.5cm]{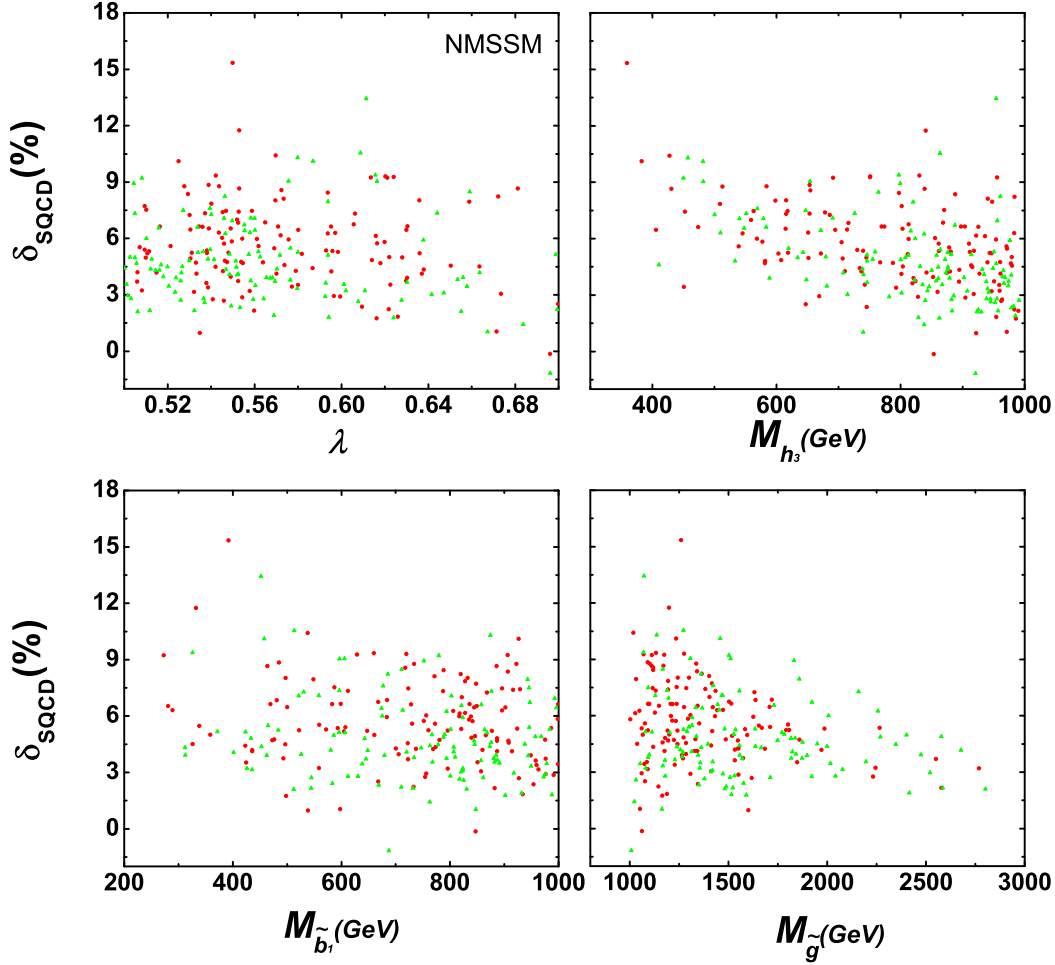}\vspace{-0.5cm}
\vspace{-1.0cm}
\caption[]{Same as Fig.\ref{fig:mssm2}, but for the NMSSM.}
\label{fig:NMSSM1}
\end{figure}

The relative SUSY-QCD corrections for the $b\bar{b} \to hh$ in the
NMSSM are presented in Fig.{\ref{fig:NMSSM1}}. It can be seen that
the maximal SUSY-QCD correction can reach $15\%$ for the samples in $1\sigma$ range. From
the upper panel we can see that $\delta_{SQCD}$ becomes small with
the increase of $\lambda$ or $m_{h_{3}}$. The reason is that with
the increase of the $\lambda$, the $m_{h_{3}}$ gets heavy and its
contribution to the cross section becomes small. From the lower
panel we see that, due to the residual effects of the sparticles,
the SUSY-QCD corrections can still reach about $9\%$ for heavy
sbottom and gluino.
\begin{figure}[ht]
\includegraphics[width=14.5cm]{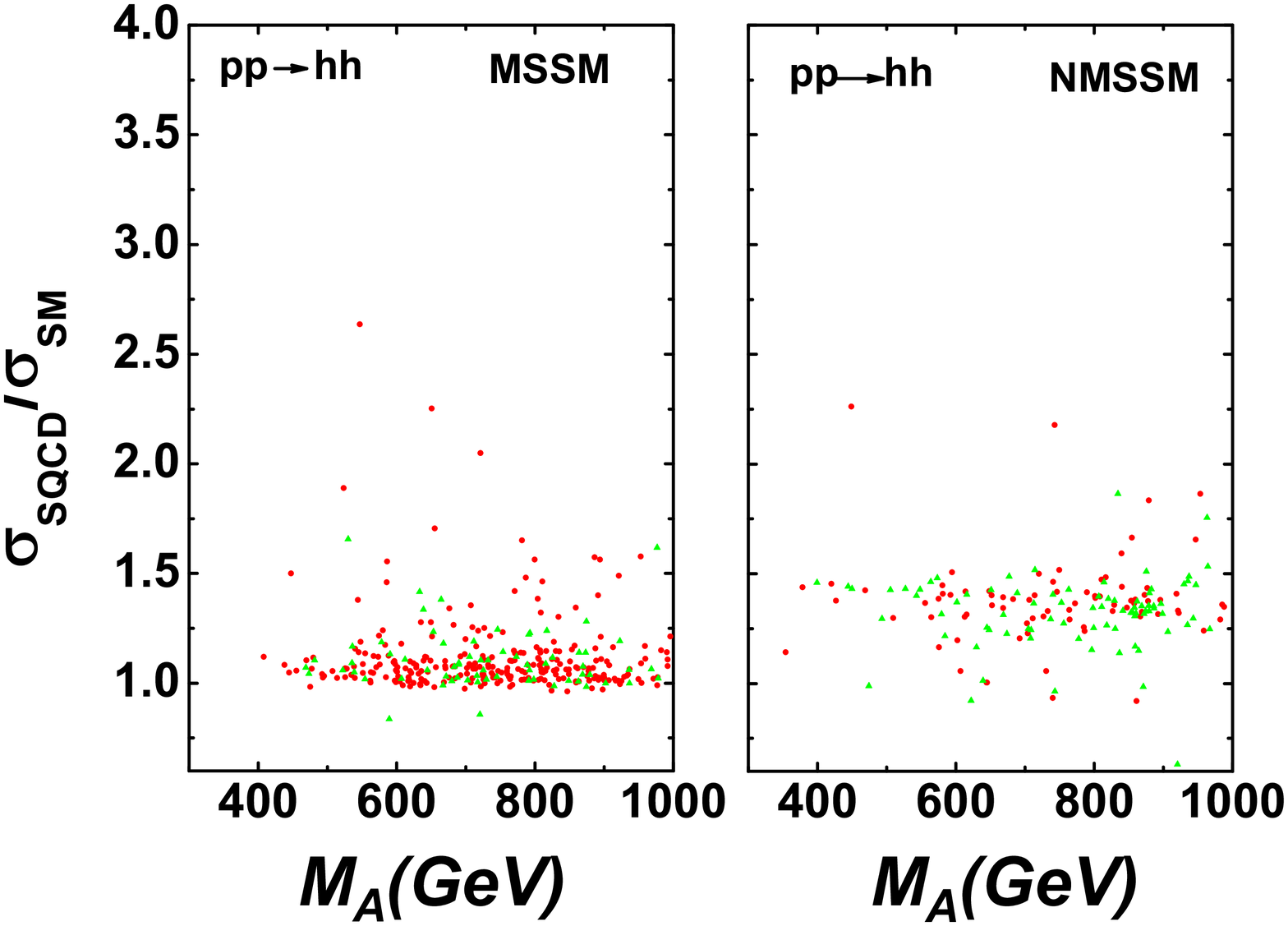}
\vspace{-0.5cm} \caption[]{The total cross section of the Higgs pair
production at the 14 TeV LHC via both $b\bar{b}$ annihilation
(include the SQCD correction) and $gg$ fusion (without the SQCD
correction) in MSSM and NMSSM.} \label{fig:combine1}
\end{figure}

In Fig.\ref{fig:combine1} we show
 the total cross section of the Higgs pair production
at the 14 TeV LHC (via both $b\bar{b}$ annihilation and $gg$ fusion)
for the samples in the $1\sigma$ and $2\sigma$ ranges of the Higgs
data. We can see that in the $1\sigma$ range the total
cross section can be maximally enhanced by a factor of 2.7 and
2.2 in the MSSM and NMSSM, respectively.

\begin{figure}[ht]
\includegraphics[width=14.5cm]{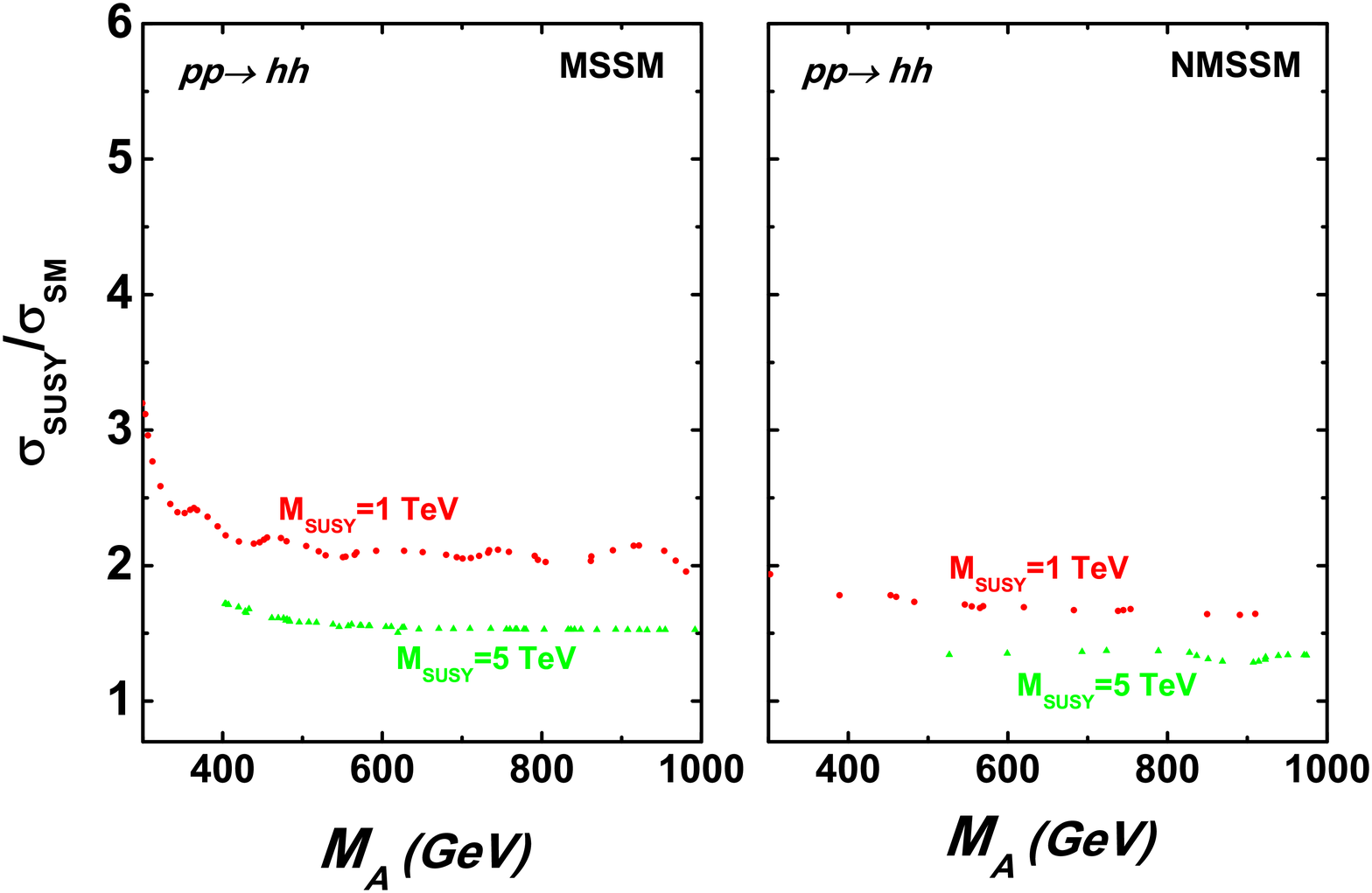}
\vspace{-0.5cm} \caption[]{The cross section of Higgs pair
production via both $b\bar{b}$ annihilation (include the SQCD
correction) and $gg$ fusion (without the SQCD correction) in MSSM
and NMSSM for heavy sparticle masses at 14 TeV LHC.
}
\label{fig:combine2}
\end{figure}

Finally, considering the null results of the direct search for
sparticles at the LHC, we investigate the SUSY-QCD effect in Higgs
pair production in the limit of heavy sparticles. For simplicity, we
assume a common mass $M_{SUSY}$ for all relevant SUSY mass
parameters:
$M_{SUSY}=M_{\tilde{Q}}=M_{\tilde{D}}=A_t=A_b=M_{\tilde{g}}=M_{\mu}$.
In Fig.\ref{fig:combine2} we display the ratio of $\sigma^{pp\to
hh}_{\rm SUSY}/ \sigma^{pp\to hh}_{\rm SM}$. We can see that for
$M_{SUSY}=1$ TeV, the ratios will maximally reach 3 and 2 in the
MSSM and NMSSM, respectively. When $M_{SUSY}$ goes up to 5 TeV, the
enhancements become weak but can still reach 1.8 and 1.4 in the MSSM
and NMSSM, respectively. So the effects of heavy sparticles decouple
quite slowly from the Higgs pair production. We checked that the
SUSY effects decouple quickly in $b\bar{b}\to hh$ but slowly in
$gg\to hh$.

\section{Conclusion}
We considered the current experimental constraints on the parameter
space of the MSSM and NMSSM.  Then in the allowed parameter space we
examined $b\bar{b}\to hh$ ($h$ is the 125 GeV SM-like Higg boson)
with one-loop SUSY QCD correction and compared it with $gg\to hh$.
We obtained the following observations: (i) For the MSSM the
production rate of $b\bar{b} \to hh$ (with one-loop SUSY QCD
correction) can reach 50 fb and thus can be competitive with $gg \to
hh$, while for the NMSSM $b\bar{b} \to hh$ has a much smaller rate
than $gg \to hh$ due to the suppression of the $hb\bar{b}$ coupling
; (ii) The SUSY-QCD correction to
$b\bar{b} \to hh$ is sizable, which can reach $45\%$ for the MSSM
and $15\%$ for the NMSSM within the $1\sigma$ region of the Higgs
data; (iii) In the heavy SUSY limit (all soft mass parameters become
heavy), the SUSY effects decouple rather slowly from the Higgs pair
production, which, for $M_{\rm SUSY}=5$ TeV and $m_A<1$ TeV, can
enhance the production rate by a factor of 1.5 and 1.3 for the MSSM
and NMSSM, respectively. Therefore, the Higgs pair production may be
helpful for unraveling the effects of heavy SUSY.

\section*{Acknowledgments}
We appreciate the helpful discussions with Junjie Cao, Ning Liu,
Wenyu Wang and Yang Zhang. This work was supported in part by
the ARC Centre of Excellence for Particle Physics at the Tera-scale,
by the National Natural Science Foundation of China (NNSFC)
under grant No. 10775039, 11075045, 11275245, 10821504 and 11135003,
by Ri-Xin Foundation of BJUT from China and by the Startup Foundation
for Doctors of Henan Normal University under contract No.11112.


\end{document}